# Magnetostructural transition in Ce(Fe$_{0.975}$Ga$_{0.025}$)$_2$ compound

Arabinda Haldar,[1] Niraj K. Singh,[2] Ya. Mudryk,[2] Ajaya K. Nayak,[1] K. G. Suresh,[1] A. K. Nigam,[3] and V. K. Pecharsky[2,4]

[1]Department of Physics, Indian Institute of Technology Bombay, Mumbai-400076, India

[2]The Ames Laboratory U. S. Department of Energy, Iowa State University, Ames, Iowa 50011-3020, USA

[3]Tata Institute of Fundamental Research, Homi Bhabha Road, Mumbai 400005, India

[4]Department of Materials Science and Engineering, Iowa State University, Ames, Iowa 50011-2300, USA

*Abstract*

The magnetic and magnetostructural properties of the polycrystalline Ce(Fe$_{0.975}$Ga$_{0.025}$)$_2$ have been investigated as a function of temperature and magnetic field. In Ce(Fe$_{0.975}$Ga$_{0.025}$)$_2$ the magnetic transition from antiferromagnetic (AFM) to ferromagnetic state (FM) is accompanied by a structural transformation from rhombohedral to cubic structure.  Phase coexistence is present during both the temperature and field driven transformations from the AFM to FM phase.

[*]Author to whom any correspondence should be addressed (E-mail: suresh@phy.iitb.ac.in).

# I. INTRODUCTION

Different classes of magnetic materials exhibiting characteristics features of the first order phase transition, such as steps in magnetization isotherms, phase co-existence, field sweep rate dependence, superheating/supercooling, unusual relaxation, etc., have attracted considerable interest recently.[1-12] Among various materials, doped $CeFe_2$ compounds have evinced the above features of the first order transition rather convincingly.[11-14] The occurrence of these exotic properties is believed to be a consequence of the magnetostructural transition.

Recently, we have reported the magnetic properties of $Ce(Fe_{1-x}Ga_x)_2$ [15]. It was shown that in $Ce(Fe_{1-x}Ga_x)_2$, the low temperature antiferromagnetic (AFM) state can be stabilized only for $x \geq 0.025$. Metamagnetic transition from the AFM to the ferromagnetic (FM) state with multiple steps was observed in the compounds with the AFM ground state. The AFM-FM transition was shown to exhibit many distinct features of first order phase transition and the martensitic scenario was invoked to explain the observed anomalous properties.[15] Therefore, in order to further improve our understanding of the Ga doped $CeFe_2$ compounds, we have carried out investigations of the magnetostructural properties of $Ce(Fe_{0.975}Ga_{0.025})_2$ using magnetization measurements and the temperature and field dependent x-ray powder diffraction (XRD) experiments. The latter technique is a unique tool to probe the structural part microscopically as a function of temperature as well as magnetic field.[16-18] Our in-field XRD investigations reveal that in $Ce(Fe_{0.975}Ga_{0.025})_2$ the magnetic and structural degrees of freedoms are intimately coupled.

## II. EXPERIMENTAL DETAILS

The polycrystalline Ce(Fe$_{0.975}$Ga$_{0.025}$)$_2$ compound was prepared as described in [15]. Magnetization measurements were carried using a commercial Physical Property Measurement System (PPMS, Quantum Design Model 6500) which has a vibrating sample magnetometer (VSM) attachment. The temperature (10-295 K) and field (0-35 kOe) dependent x-ray powder diffraction data were collected using a Rigaku TTRAX powder diffractometer with Mo $K_\alpha$ radiation.[19] Multiple sets of diffraction data were collected in step scanning mode (0.5–2 s/step) with a 0.01° step of $2\theta$ over the range of $9° \leq 2\theta \leq 45°$. Each data set was analyzed by the Rietveld refinement to determine the unit cell parameters and the phase contents, when two different crystallographic phases coexisted in certain field and temperature regimes.

## III. RESULTS AND DISCUSSION

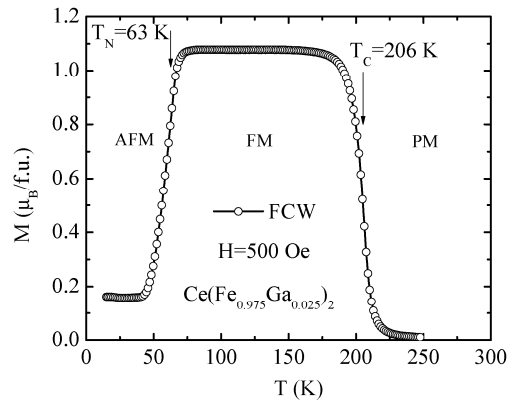

FIG. 1. Temperature variation of magnetization of Ce(Fe$_{0.975}$Ga$_{0.025}$)$_2$ compound in field cooled warming (FCW) mode under an applied field (*H*) of 500 Oe.

Figure 1 shows the temperature variation of magnetization (*M*) data of Ce(Fe$_{0.975}$Ga$_{0.025}$)$_2$ measured in field cooled warming (FCW) mode under an applied field (*H*) of 500 Oe. We note that owing to the presence of AFM state the magnetization at low temperatures is quite small. On warming the compound undergoes two magnetic transitions: AFM to FM state and then from FM state to paramagnetic (PM) state. These transitions occur at 63 K ($T_N$) and 206 K ($T_C$).

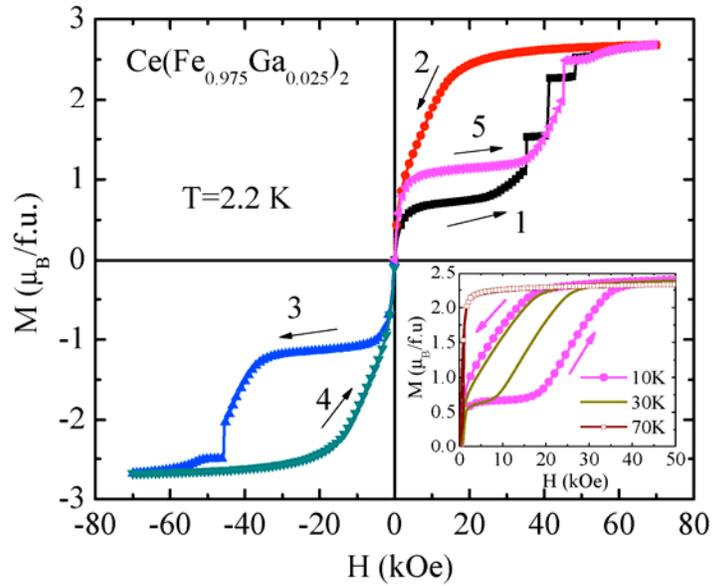

FIG. 2. Magnetization isotherm of Ce(Fe$_{0.975}$Ga$_{0.025}$)$_2$ at T=2.2 K. Inset shows the isotherms at higher temperatures.

Selected magnetization isotherms of Ce(Fe$_{0.975}$Ga$_{0.025}$)$_2$ are shown in Fig. 2. We note that in the AFM regime, the isotherms show multiple steps, which are attributed to the field induced AFM - FM transition. The M(H) data also show large hysteresis and the virgin curve (labeled as 1) lies outside the envelope curve (labeled as 5). These features are attributed to the supercooling/superheating effect and the kinetic arrest of first order phase transition.[15] A similar step behavior in magnetization isotherms has been reported in manganites[1-8] and in a few intermetallic compounds[10,11]. Similarities in the magnetization behavior in different classes of materials and a universal picture arising from their study have been reported by Roy et al.[20] We believe that a martensitic scenario arising from the structural mismatch between the FM and the AFM phases leads to the observed step behavior in this system.[15]

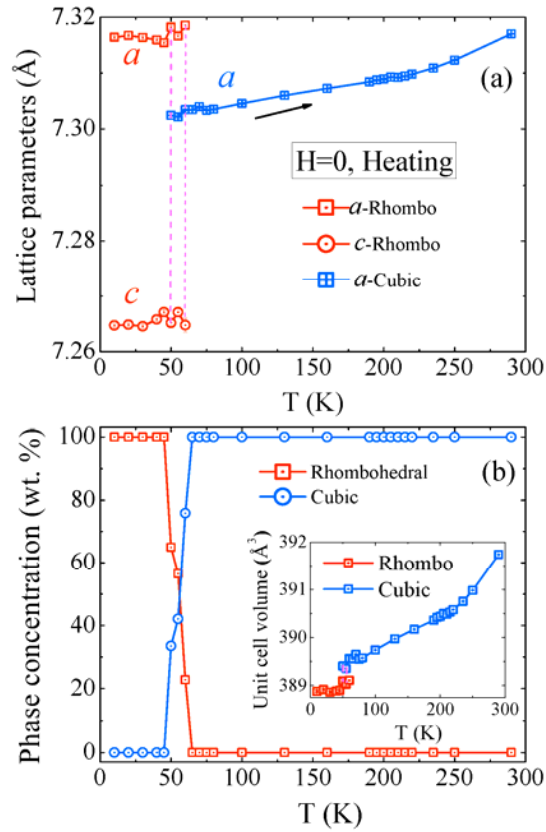

FIG. 3. (a) Temperature variation of the normalized lattice parameters of Ce(Fe$_{0.975}$Ga$_{0.025}$)$_2$ during heating in zero field. (b) Temperature dependencies of the concentrations of the rhombohedral and cubic phases obtained from zero field x-ray diffraction data. Inset shows the variation of unit cell volume with temperature

It was mentioned above that a martensitic scenario may be responsible for the anomalous magnetic properties of Ce(Fe$_{0.975}$Ga$_{0.025}$)$_2$. Thus, in order to further understand the Ce(Fe$_{0.975}$Ga$_{0.025}$)$_2$, we have carried out the temperature dependent x-ray powder diffraction measurements and the results of Rietveld refinement of XRD results are shown in Figure 3. Here, the sample was zero field cooled to 10 K, and XRD data were collected during warming. At 10 K, the compound possesses rhombohedral structure, which is preserved up to 45 K. With increase in temperature, the compound transforms from the rhombohedral to the cubic structure and, between 45 and 65 K, both polymorphs coexist. At 65 K the compound completely adopts the cubic structure and this structure is retained up to room temperature. We note that the lattice parameters of the rhombohedral phase shown in Fig. 3a were calculated by the Rietveld refinement using hexagonal setting of the R3m space group and the calculated $a_{rh.}$ and $c_{rh.}$ parameters were modified ($a_{rh.}$ was multiplied by $\sqrt{2}$, and $c_{rh.}$ was divided by $\sqrt{3}$) in order to be directly comparable with the high-temperature cubic lattice parameter. The unit cell volume (Inset of Fig. 3b) of the rhombohedral structure was also normalized (V = $V_{rh.} \times 4/3$).

Therefore, the XRD data reveal that in the low temperatures AFM phase, Ce(Fe$_{0.975}$Ga$_{0.025}$)$_2$ possesses the rhombohedral structure whereas in the high temperature FM and PM phases it adopts the cubic structure. The same is clearly illustrated in Fig. 3b by calculated from x-ray data phase fractions of different phases as functions of temperature. We note that the temperature range over which the rhombohedral to cubic transformation occurs (45-65 K, see Fig. 3) matches well with the broad AFM-FM transition seen in the *M(T)* data of Fig. 1.

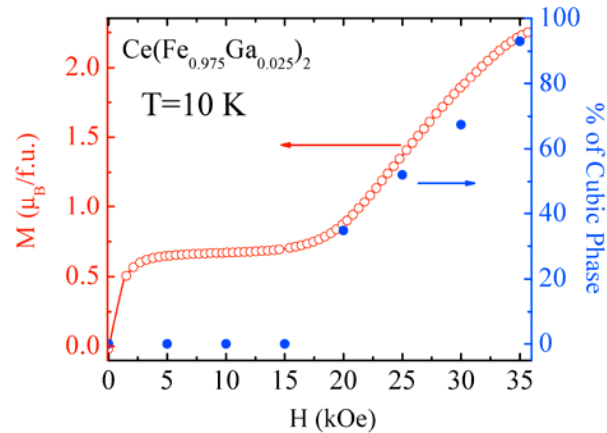

FIG. 4. Field ($H$) dependence of magnetization ($M$) and the percentage of cubic phase in Ce(Fe$_{0.975}$Ga$_{0.025}$)$_2$ at T = 10 K.

Figure 4 shows the results of the field dependent x-ray powder diffraction examination carried out at 10 K along with the M(H) isotherm of Ce(Fe$_{0.975}$Ga$_{0.025}$)$_2$ measured at the same temperature. Here the sample was zero field cooled to 10 K and then the field dependent XRD patterns were recorded with a field step of 5 kOe up to a maximum field of 35 kOe. The field dependent XRD data reveal that in zero field the Ce(Fe$_{0.975}$Ga$_{0.025}$)$_2$ possesses rhombohedral structure and this structure is preserved up to H = 15 kOe. However, at H = 20 kOe, about 35% of the rhombohedral phase is converted to the cubic phase. It is interesting to note that the ZFC M(H) of Ce(Fe$_{0.975}$Ga$_{0.025}$)$_2$ also shows the metamagnetic phase transition at the same field. With further increase in the field the concentration of the cubic phase increases and at 35 kOe field, about 93% of the rhombohedral phase is converted into the cubic phase. Thus, the field dependent XRD data reveal that in Ce(Fe$_{0.975}$Ga$_{0.025}$)$_2$ the metamagnetic transition from AFM to FM state is associated with the field induced structural transformation from rhombohedral to cubic phase. We find that the growth of the cubic phase follows the evolution of the ferromagnetic phase (see Fig 4). Thus, both the temperature and field dependent XRD results indicate that in Ce(Fe$_{0.975}$Ga$_{0.025}$)$_2$ the magnetic and structural degrees of freedoms are intimately coupled.

There are many similarities between the properties of Ce(Fe$_{0.975}$Ga$_{0.025}$)$_2$ and other materials which exhibit first order magnetostructural transition.[16-18] Dependence of measurement protocols, like magnetic field sweep rate, time delay, etc. was reported earlier for the title material.[15] The underlying physics of these phenomena can be explained with this evidence of structural transition along with the magnetic transition. The slow relaxation observed in this compound arises due to the mismatch between experimental time scale and the transformation time scale of the lattice structure.[21] So the experimental time scale produces metastable states which relax to equilibrium state when the specimen is allowed to relax. As the moment and the lattice structure are strongly coupled with each other, jumps are expected in magnetization isotherms as well as in magnetoresistance.

**CONCLUSION**

Temperature and field variation of structural properties of Ce(Fe$_{0.975}$Ga$_{0.025}$)$_2$ have been studied using x-ray powder diffraction technique. The strong magnetostructural coupling seen from the combination of the XRD and magnetization data lends direct experimental evidence to the martensitic scenario predicted in this compound. The observation of the first order nature of the magnetostructural transition between the AFM and the FM states and the phase co-existence of structural and the magnetic phases over a certain regime of temperature and field are the highlights of this study. Comparison with the different classes of materials (magnetocaloric materials and magnetoresistive oxides) brings to light the universality of phase separated systems.

**ACKNOWLEDGEMENTS**

KGS and AKN thank BRNS (DAE) for the financial assistance for carrying out this work. The Ames Laboratory is supported by the Office of Basic Energy Sciences, Materials Sciences Division of the U.S. Department of Energy under contract No. DE-AC02-07CH11358 with Iowa State University. The authors thank Prof. K.A. Gschneidner, Jr. for fruitful discussions.